\documentclass[manuscript]{aastex}
\usepackage{graphicx}
\usepackage{amsmath}

\shorttitle{Planetary Systems in Binary Stars: The Location of Secular
Resonances}
\shortauthors{Pilat-Lohinger et al.}

\begin{document}

\title{A quick method to identify secular resonances in multi-planet systems
  with a binary companion}

\author{E. Pilat-Lohinger, A. Bazs\'o}
\and
\author{B. Funk}

\affil{Institute for Astronomy, University of Vienna, 
      T\"urkenschanzstrasse 17, A-1180 Vienna, Austria}

\begin{abstract}
Gravitational perturbations in multi-planet systems caused by an
  accompanying star are the subject of
  this investigation. Our dynamical model is based on the
  binary star HD41004\,AB where a giant planet orbits HD41004\,A. We modify
  the orbital parameters of this system and analyze 
  the motion of a hypothetical test-planet surrounding HD41004\,A on an
  interior orbit to the detected giant planet. Our numerical computations
  indicate perturbations due to  mean
  motion and secular resonances. The locations of these resonances are usually 
  connected to high eccentricity and highly inclined motion depending strongly 
  on the binary-planet architecture. 
  As the positions of mean motion resonances can easily be determined,  the main
  purpose of this study is to present a new semi-analytical method to
  determine the location of a secular resonance without huge computational effort.
\end{abstract}


\section{Introduction}

Detections of planets in binary star
systems \citep[see e.g.][]{roell12} and the 
knowledge that such stellar systems are quite numerous in the solar
neighborhood \citep{ducma91, rag10, tok14} have
encouraged astronomers to  
study the planetary motion in such systems. 
In this context, we have to distinguish different types of motion
\citep{dvo84}:  
 (i) the circumstellar 
or S-type orbits where the planet moves around one stellar component and
(ii) the circumbinary or P-type motion where the planetary orbit surrounds both
stars. For completeness we mention a third type, known as L- or
  T-type (or Trojan motion),
where the planet moves in the same orbit as the secondary star but $60^o$
ahead or behind the secondary. A restriction in the mass-ratio makes this motion
less interesting for binary stars. \\
Even though it was questionable that planets could exist in
binary star systems (especially in tight systems), scientists working in
dynamical astronomy  have shown very early that planetary motion in binary star systems could be
possible in spite of the gravitational perturbations of the second star. 
Such studies have been carried out long before the detection of planets
outside the solar system \citep[see e.g.][]{har77, grabl81, bla82,
dvo84, dvo86, radvo88, dvo+89} and various studies
by Benest between 1988 and 1993.
The discoveries of a planet in the $\gamma$ Cephei system \citep{coc+02} and
in the Gliese 86 system 
\citep{quel+00}, respectively led to a re-processing of such stability
studies \citep{HW99, pildvo02, pilo+03}.
Meanwhile about 80 binary star systems are known to host one or several 
planets\footnote{see 
  http://www.univie.ac.at/adg/schwarz/multiple.html (A binary catalogue of
  exo-planets maintained by R.\ Schwarz).} and most of them are in
circumstellar motion. \\
However, the studies cited above provide only stability limits for a
single planet in a binary star system.
In this study, we investigate the circumstellar motion of two planets  and
analyze the perturbations caused by the secondary star.  Our investigation is
based on the work by \cite{pilo05} where the influence of a secondary star 
on two-planets has been studied for the two
spectroscopic binaries $\gamma$ Cephei and HD41004\,AB. In both systems,  the
secondary star is a M-dwarf in a distance of about 20 au from 
the host-star of the detected gas giant.
The numerical study by  \cite{pilo05} showed significant deviations in the
stability maps of a certain area in the two binary systems due to different
masses of the host-stars 
and the discovered planets and the different positions of the latter.
A variation of the giant-planet's semi-major axis provided a first explanation
for these deviations.  \\
In the present investigation, we studied the locations of gravitational
perturbations\footnote{We did not study  planetary migration in this 
investigation since we assumed that the formation process was already
completed.} like mean motion resonances (MMRs)  
and secular resonances (SRs) in the binary system HD41004\,AB
using numerical computations, taking into account different binary-planet
configurations for which we varied
(i) the mass of the secondary star and (ii) the eccentricities of the binary star
and the detected giant planet. For these configurations, we calculated
  stability maps for the motion of 
test-planets orbiting the host-star interior to the giant planet. The
numerical results led us to study the planetary motion in detail using
a frequency analysis for planar orbits. 
 Moreover, as an experiment we calculated the proper frequencies of the
  test-planets using the Laplace-Lagrange secular perturbation theory
which resulted in  a new semi-analytical approach that
permits a fast determination of the location of a linear SR.
The SR is a striking feature in many dynamical maps in our study and might be
important for habitability studies of such systems.\\ 
This article is structured as follows: in section 2, we explain the
dynamical model and initial conditions for the computations and we discuss
the perturbations on the planetary motion. 
Section 3 shows the numerical results for different binary-planet
configurations for which we analyze the behavior and the perturbations on the
motion 
of the test-planets. In section 4, we present a new semi-analytical
approach to 
determine the location of the secular perturbation and an
application of our method to different
binary stars-planet configurations of HD41004\,AB. Then we compare the results
with those of the numerical study and finally, in section 5 we summarize our study.

\section{Numerical computations}

\begin{table}[h]
  \begin{center}
  \caption{Initial conditions for the computations (heliocentric -- i.e.\ with
    respect to $m_1$):}
  \vspace{5mm}
  \label{tab1}
 {\scriptsize
  \begin{tabular}{|r|c|c|c|}\hline 
{\bf masses: $m_1$, $m_2$ [$m_{Sun}$]}& {\bf semi-major axis($a$)} & {\bf
  eccentricity ($e$) }& 
{\bf inclination ($i$), argument of perihelion ($\omega$) }\\ 
{\bf  $m_3$, $m_4$ [$m_{Jup}$]}  & {\bf [AU]} & & {\bf node ($\Omega$), mean anomaly ($M$) } \\ 
\hline
$m_1 = 0.7$ & & & \\
\hline
$m_2 = 0.4 / 0.7 / 1.0 / 1.3 $ & 10 -- 50 & 0.0 -- 0.6 &  $i_2, \omega_2,
\Omega_2, M_2 = 0$ \\ 
& $\Delta a = 10$ & $\Delta e =0.2$ & \\
\hline
$m_3 = 2.5$ & 1.64 & 0.2, 0.4 & $i_3, \omega_3, \Omega_3, M_3 = 0$ \\
\hline
$m_4 = 0.0$ & 0.15 -- 1.3 & 0.0 & $i_4, \omega_4, \Omega_4, M_4 = 0$ \\
\hline
 \end{tabular}
  }
 \end{center}
\vspace{1mm}
\end{table}  

For our numerical study of circumstellar motion of two planets in a binary
star system we used a configuration resembling the
HD41004\,AB system where a giant planet has been discovered 
\citep{zuck+03,zuck+04}. This tight binary star is about 43 pc away from our 
sun. Both stellar components, a K2V and a M2V star, are accompanied by either 
a giant planet or a brown dwarf, where the latter is ignored in our
dynamical model. The parameters of the two stars were taken from the paper
by \cite{roell12}, where the two stellar masses are 0.7 M$_{Sun}$
and 0.42 M$_{Sun}$, respectively.
While the distance of the two stars seems to be well determined with about 20
au, the binary's eccentricity is not known\footnote{In this
  context, \cite{pilfun10} studied the stability of this binary system using 
different published values of the eccentricity of HD41004\,Ab to get upper
limits for the binary's eccentricity.} and needs further observations.
The detected giant planet surrounds HD41004A at 1.64 au, well inside
the stable area for $e_B=0.4$ is 2.8 au\footnote{The published value for
  $e_B=0.4$ is 3.38 au \citep{roell12}  
which is larger as they considered circular motion of the test-planets
\cite[according to][]{HW99}.}.
And the additional hypothetical test-planet was placed internal to
the giant
planet in an initially circular orbit. In order to save computation time
  we used the  
restricted four body problem where the test-planet has a negligible mass
compared to the binary stars and the gas giant. However, 
test-computations with massive test-planets (up to an Earth-mass) showed the
same dynamical behavior.
The equations
of motion were calculated numerically by means of the Lie series method using
the {\it nine package} of Eggl \citep[see][]{egg-dvo10}. Table \ref{tab1} summarizes  the
different initial binary-planet configurations which were calculated for a
time span between $10^6$ and $5 \times 10^6$ years.

\subsection{Influence of the secondary star}

\begin{figure}
\centering
\includegraphics[scale=0.6]{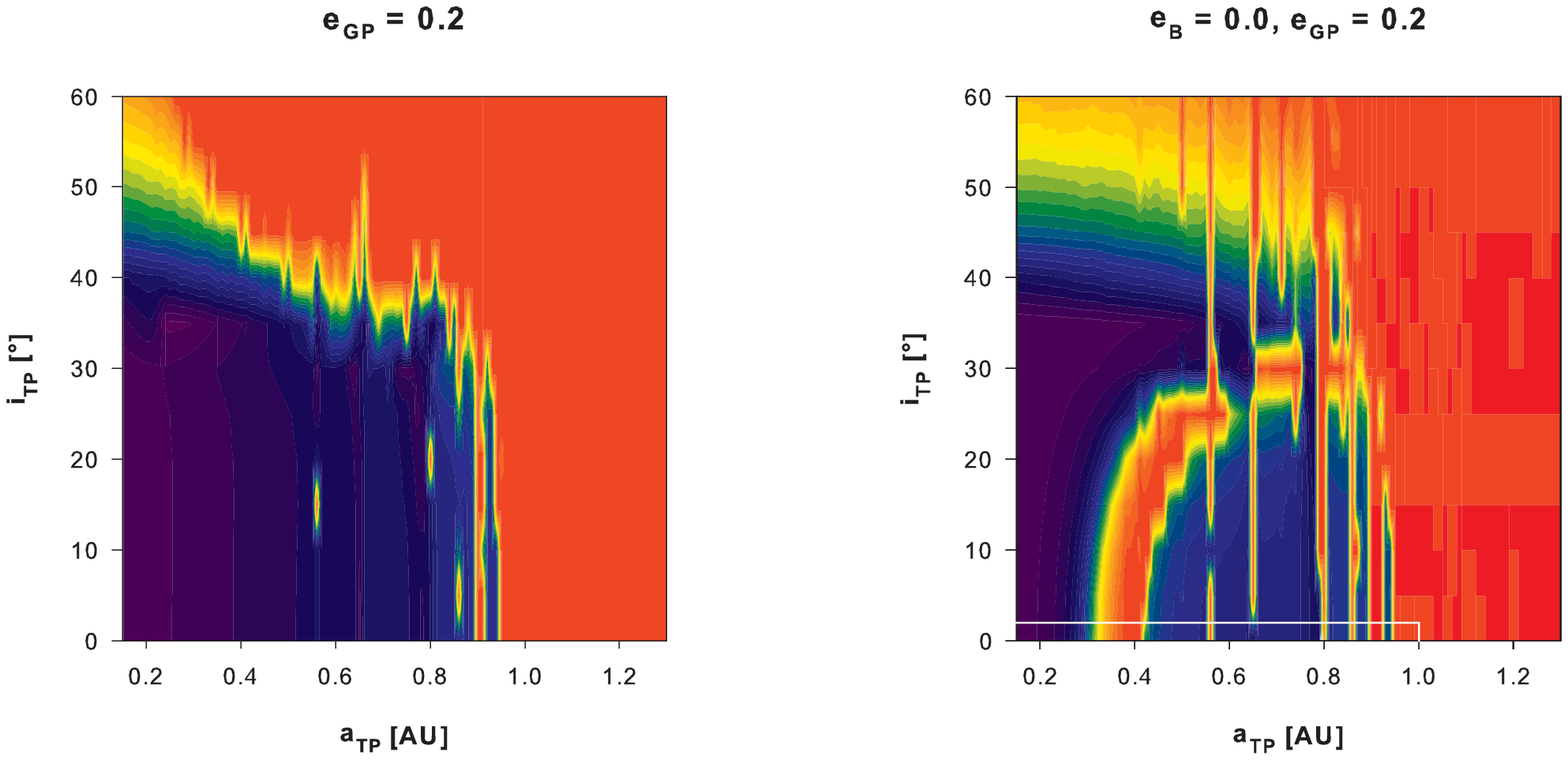}
\includegraphics[]{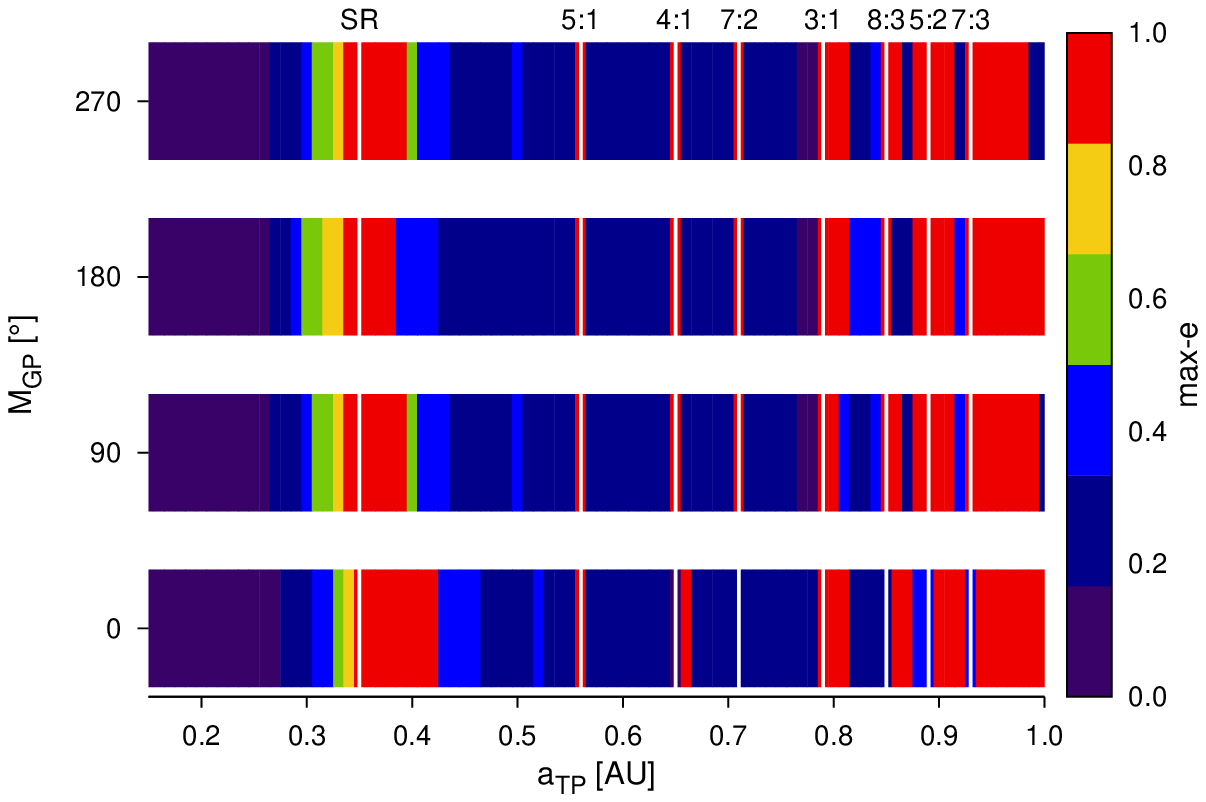}
\caption{Max-e plots for (a) test-planets perturbed
  by a giant planet at 1.64 au (upper left panel) and (b) test-planets perturbed by
  the giant planet and a secondary (M-type) star at 20 au (upper
  right panel). The lower plot shows the max-e values for the area marked by the white
  rectangle in the upper right figure for 
  various initial mean anomalies of the giant planet ($M_{GP}$) (see y-axis).
The color code (which is the same in all max-e maps) indicates the different
maximum 
eccentricities achieved by the test-planets over the whole computation time. Red
labels the unstable area and purple indicates  circular motion. 
\label{F1}}
\end{figure}

In a hierarchical four body problem gravitational perturbations like
mean  motion resonances (MMRs) and secular resonances (SRs) influence the
architecture of a planetary system. Such perturbations are well visible in the
various maps (Figs.\ 1, 2, 4, 5) which display the maximum eccentricity
(max-e) of test-planets orbiting HD41004\,A in the region between 0.15 and 1.3
au for various inclinations. The color code defines regions of different
max-e values (from 0 to 
1), where red marks the unstable zone.
Most of these maps show vertical straight lines indicating MMRs. In
addition, a strong perturbation is visible by an arched red band within the
stable area. We will show in section 4 that this perturbation is a linear SR.
In certain maps the SR is not visible (see e.g.\ Figs.~\ref{F4} right panels)
because it is located either in the unstable (red) area or too close to the
host-star.  \\
The upper left panel of Fig.~\ref{F1}, does not show the SR as it represents the
dynamics in the restricted three body problem where the secondary star was not
taken into account. This plot indicates well known dynamical features of 
the restricted three body problem, like 
MMRs\footnote{See also table \ref{tab2}.} with respect to the giant planet
(shown by the yellow spikes) and the cut-off of the stable region for
inclinations $ > 38^o$ caused by the so-called Kozai resonance \cite{koz62}. \\
As soon as we add the secondary star at 20 au to the system a red
arched band appears in the stable area (see Fig.\ 1 upper right
panel) which arises from a perturbation of the  giant planet
and the secondary star (for details see the forthcoming section). \\
A comparison of the two upper panels of Fig.~\ref{F1} shows that the extension of the
stable area in $a_{TP}$ is about the same which indicates
that the border between stable and unstable motion is
determined by the giant planet's semi-major axis. The main differences of
circumstellar motion in a binary system and the motion around a single star
are the following: \\ 
(i) In addition to MMRs a SR might appear under certain circumstances. \\
(ii) The MMRs with respect to the giant planet are stronger and therefore
better visible as the giant planet is perturbed by the secondary star leading
to a fluctuation in $e_{GP}$.\\
(iii) The border of unstable motion is shifted to higher
inclinations ($i_{TP} > 60^o$) of the test-planets. 

\begin{table}
  \caption{MMRs in the  HD41004\,A system for $a_{GP}=1.64$ AU:}
  \begin{center}
\label{tab2}
 {\scriptsize
  \begin{tabular}{|l|c|c|c|c|c|c|c|c|c|c|c|c|c|c|}\hline 
MMR & 2:1&3:1& 4:1&5:1&5:2&5:3&6:1&7:1&7:2&7:3&7:4&8:3&9:2&9:4 \\
\hline
Position[AU]& 1.03&0.79&0.65&0.56&0.89&1.16&0.49&0.45&0.71&0.93&1.13&0.85&0.6&0.95 \\
\hline
 \end{tabular}
  }
 \end{center}
\end{table}

Moreover, the right panel of Fig.~\ref{F1} shows  that the
region  between the host-star and the SR is obviously not perturbed by the giant
planet as the motion of the test-planets remains nearly circular during the
whole computation.   So we assume that this area would provide
best conditions for {\it dynamical habitability} as nearly circular
planetary orbits will most probably be fully in the so-called habitable zone
(HZ)\footnote{The HZ is the region around a star where a terrestrial planet
  would have appropriate conditions for the evolution of life.}. \\
The white rectangle in this plot marks the planar motion of the test-planets
for which we varied the initial mean anomaly of the giant  ($M_{GP}$) and of the
test-planet ($M_{TP}$) where the variation of the latter yield same results. 
The dynamical behavior in the labeled area for various $M_{GP}$ is shown in
the lower panel of Fig.~\ref{F1}. One can see that most of the resonances
(indicated by the white vertical lines)
appear even if we change the initial relative positions of the two
planets. The 7:2 and the 5:2 MMRs are examples that are not visible for all
calculated $M_{GP}$ values.

\section{Results of the numerical study}
\begin{figure}
\centering
\includegraphics[scale=0.6]{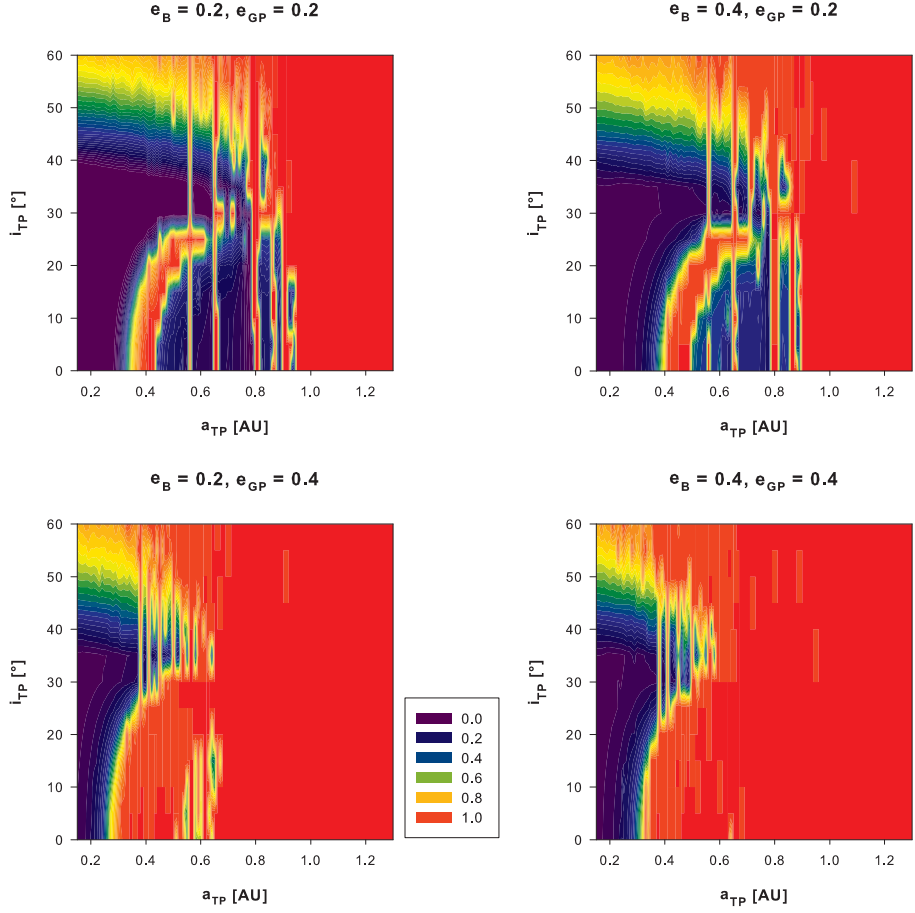}
\caption{Max-e maps for planetary motion in the binary HD41004
  where $a_B = 20$ au in all panels. Only the  initial values of $e_B$ and
  $e_{GP}$ are changed between 0.2 and 0.4 as indicated in the title 
  of each panel. The color code is the same as in Figs.\ 1.}
\label{F2}
\end{figure}

The initial conditions shown in table \ref{tab1} yield in total 320 different
binary-planet configurations for which we
calculated max-e maps consisting of 1560 orbits each in the ($a_{TP}, i_{TP}$)
plane. Here $a_{TP}=0.15 \ldots 1.3$ au and $i_{TP}=0^o \ldots 60^o$ denotes
the semi-major axis and the 
inclination of the test-planet as shown in Figs.~\ref{F2}a-d. The color code
indicates 
 the different maximum eccentricities of the test-planets: from circular
(purple) to highly eccentric motion (yellow). The unstable orbits are labeled
by the red area. The four figures
show the results for the same distance of the two stars ($a_B = 20$ au) and
the same mass of the secondary star ($m_2= 0.4 m_{Sun}$). Only the initial
eccentricities of the binary $e_B$ and of the giant planet $e_{GP}$ were
varied between 0.2 and 0.4 for both.

\subsection{Effect of $e_B$ and $e_{GP}$ on the dynamics of the test-planets}
\begin{figure}[h]
\centering
\includegraphics[scale=0.45]{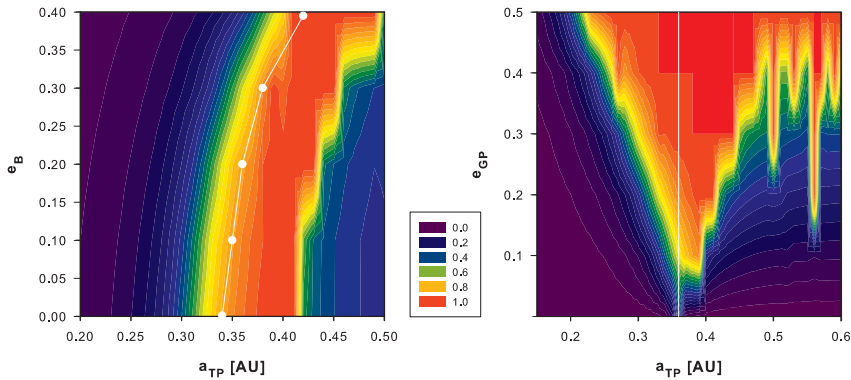}
\caption{Max-e maps for HD41004\,AB with $a_B=20$ au. The
  left panel shows for a fixed value of $e_{GP} (=0.1)$ the displacement of the
  perturbed area (red) when $e_B$ (y-axis) is increased. Locations of the SR
  derived with our method are given by the white line of points.
  The right panel shows for a fixed $e_B (=0.2)$ an
  enlargement of the perturbed area (red) when $e_{GP}$ (y-axis) is
  increased, while the vertical white line represents the location of the SR
  calculated with our method.}
\label{F3}
\end{figure}

A comparison of the different panels of Fig.~\ref{F2}(a-d) shows clearly that
the planet's eccentricity affects the stable area stronger than the binary's
eccentricity. An increase of $e_{GP}$ shrinks the stable zone significantly,
which can be seen when we compare either the two left panels or the two right panels
of Fig.~\ref{F2}. Especially the area to the right of the
arched red band is severely perturbed if
$e_{GP}$ is increased from 0.2 to 0.4. In the lower left panel of Fig.~
\ref{F2}, one can see that the area of the SR is enlarged due to
the higher $e_{GP}$. An increase in $e_B$ from 0.2 to 0.4 indicates only a
slight shift of this perturbation  to the right (compare the two upper
panels of Fig.~\ref{F2}). \\
Due to these changes, we have plotted in Figs.~\ref{F3} the extension of this perturbation
for different eccentricities of the binary (left
panel) and of the giant planet (right panel). These two panels confirm that
changes in $e_B$ will modify the location of the SR while 
an increase of $e_{GP}$ enlarges the perturbed (red) area  which is visualized
in the 
right panel of Fig.~\ref{F3} showing a V-shape for this perturbation. This
phenomenon is well known for MMRs in such ($a,e$) maps.  

\subsection{Effect of $a_B$ on the dynamics of the test-planets}
\begin{figure}
\centering
\includegraphics[scale=0.6]{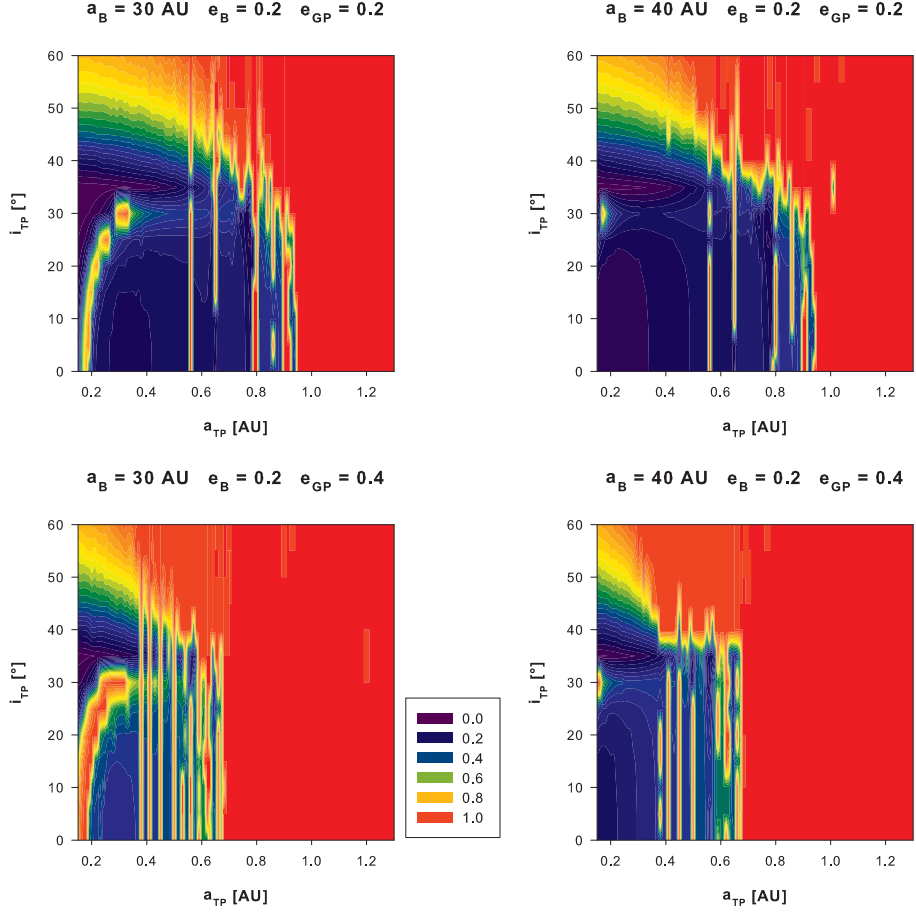}
\caption{Max-e maps for different distances of the two stars
  and different $e_{GP}$, while $e_B$ is fixed to 0.2.
Left panels show the perturbations for the binary with $a_B = 30$ au and the
right panels show the same for $a_B = 40$ au. }  
\label{F4}
\end{figure}
A variation of the distance of the two stars ($a_B$) and the
influence on the orbital behavior of the test-planets is given in 
Figs.~\ref{F4}(a-d). We show the results for two stellar separations: 30 au 
(left panels) and 40 au (right panels). The eccentricity of the binary is
fixed to 0.2 and $e_{GP}$ is either 0.2 (upper panels) or 0.4 (lower panels),
respectively. A comparison with the equivalent results for $a=20$ au (see
Fig.~\ref{F2}, left panels) shows a significant shift of the SR
towards the host-star. For $a_B= 30$ au, the arched red band is
still visible but significantly reduced in the width. Moreover, in the upper
left panel of Fig.~\ref{F2} the color of the SR has
changed to yellow with only a few red spots. This means that
the orbits in this area are more stable but with strong periodic
variations in eccentricity (up to $e=0.8$). An increase in $e_{GP}$  (see lower
left panel) shows the following
effects: (i) a significantly smaller stable
region, (ii) an enhancement of the MMRs, (iii) an enlargement
of the area of the SR and (iv) stronger perturbations within
the SR. \\
A further increase of the stellar distance causes again a shift of the SR
towards the host-star so that it moves out of the area we are
studying. Only the yellow/red spot at $i_{TP}=30^o$ indicates the existence
of this phenomenon (see right panels of Fig.~\ref{F4}). A SR
close to the host-star could cause periodic variations in
eccentricity for close-in planets. How tidal effects might influence
the orbital behavior in addition could be an interesting aspect for a future
study. \\ 
Figure \ref{F4}.d (lower right panel) shows another new feature, namely the
sharp border between stable and unstable motion at $i_{TP}=38^o$
for the area perturbed by MMRs with respect to the giant planet (for $a_{TP}$
from 0.4 to  
nearly 0.7 au). As this feature can be observed only for a low-mass secondary
star we assume that in such a system this area is only affected
gravitationally by the gas giant. 
\subsection{Effect of the secondary's mass on the dynamics of the
  test-planets} 

\begin{figure}[h]
\centering
\includegraphics[scale=0.45]{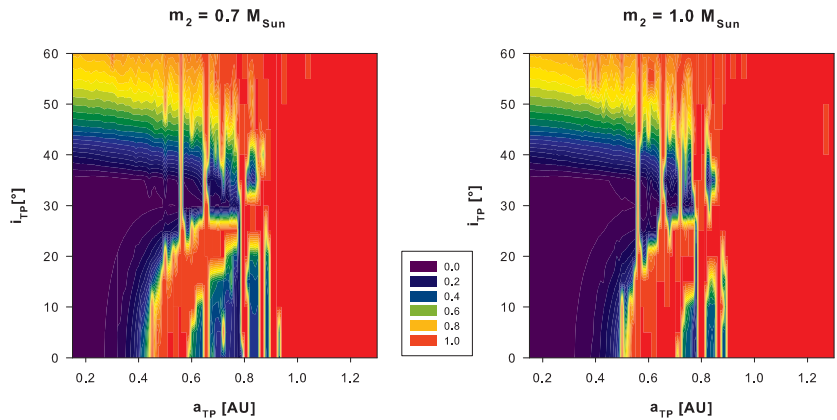}
\caption{Max-e maps for test-planets around HD41004A perturbed
  by the giant planet at 1.64 au and the secondary star at 20 au. In the left
  panel, the secondary is a K-type star of 0.7 solar masses. In the right
  panel, the secondary is a G2-type star of 1 solar mass. The initial
  eccentricities are 0.2 for all massive bodies.}
\label{F5}
\end{figure}
The influence of the mass of the secondary star is shown in Figs.~\ref{F5}(a,b).
The two panels display the result of the same configuration as for the study 
of  Fig.~\ref{F2} upper left
panel only the  mass of the perturbing star at 20 au is increased. A comparison of
these plots shows a shift of the SR to larger semi-major
axes, i.e. in direction towards the giant planet and the secondary star. In
addition, we recognize an enlargement of the perturbed area due to the fact
that a more massive secondary star has a stronger influence on the giant
planet, so that the eccentricity of the giant planet is increased and the
perturbed area will grow simultaneously. \\

\section{Semi-analytical approach to locate a linear secular resonance}

\begin{figure} 
\centering
\includegraphics[scale=.9]{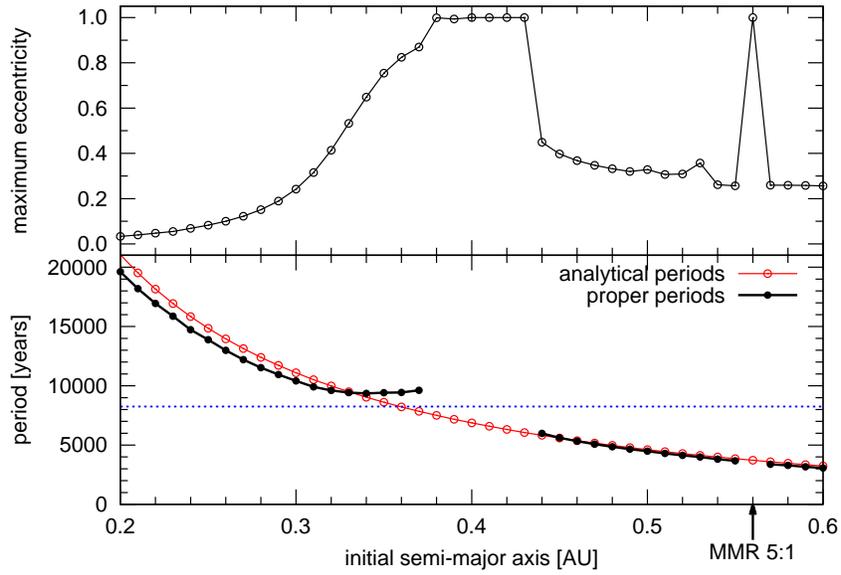} 
\caption{The upper part shows the maximum eccentricity of test-planets in
  planar motion in
  the region between 0.2 and 0.6 au. The lower part
  indicates the proper periods of the test-planets: black dots are the
  numerically determined values using a FFT method and red dots
  represent the analytical solution. The horizontal blue line indicates the
  proper period of the giant planet.}
\label{Fpf}
\end{figure}

To explain the strong perturbation represented by the arched red band in various
max-e maps (see e.g.\ Fig.~\ref{F2}a-d) we studied in detail the area 
between 0.2 and 0.6 au around HD41004A. In Fig.~\ref{Fpf}
(upper panel), we show
the maximum eccentricity of the test-planets in planar motion in this
region. Perturbations are indicated by higher values of max-e
which is the case for the orbits at 0.53 au and 0.56 au and for all orbits in the area
between 0.3 and 0.44 au. For this region, the 
lower part of Fig.~\ref{Fpf} shows that the  proper periods of
the test planets are similar to the proper period of the giant planet which is
given by the horizontal blue line in this panel. Consequently, this perturbed
area is due to a linear secular resonance (1:1) where the frequencies of
precession of the orientation of the orbits in space of the test-planets and
of the giant planet are the same . \\ 

\subsection{Numerical values for proper frequencies}

To determine the proper frequencies of the orbits one needs
long-term computations of the dynamical system where the integration time depends on
the distance of the two stars, which is in our case about 20 au. For such a
tight binary, we
needed only calculations over some $10^6$ years. Of course the computation time
increases with $a_B$ \citep[as discussed in][]{baz+16}. \\
The time evolution of the orbits was analyzed using the Fast Fourier
Transform library FFTW of \cite{fri-joh05}\footnote{see http://fftw.org/}.
In addition, we applied the tool {\it SigSpec} of \cite{reg07} for our orbital
analysis where a Discrete Fourier Transform is used to decide whether a peak
in the Fourier spectrum is not due to the noise in the signal. From the
main frequency in the Fourier spectrum of an orbit we calculated the proper
period. The results are shown by the black dots in the lower panel of
Fig.~\ref{Fpf}. It is clearly seen that we could not determine the proper
periods for orbits in the area between 0.38 and 0.43 au  and at 0.56 au,
i.e.\ for orbits with max-e$=1$ (compare the two panels
of Fig.~\ref{Fpf}). As the Fourier spectrum indicates a
chaotic behavior for these orbits we could not determine the proper
frequencies. \\ 
In addition we tested the application of a secular perturbation theory for the
determination of the proper frequencies of the test-planets.

\subsection{Analytical values for proper frequencies}

The well-known Laplace-Lagrange
secular perturbations theory \citep[see e.g.][]{mur-der99} was used to derive
an analytical solution for the proper frequencies.
As this theory has been developed for studies in the solar system, it is restricted to  
low eccentricities and low inclinations. Nevertheless, we tested its
application for the HD41004 binary system using an eccentricity of 
0.2 for both, the binary and the giant planet.
The massless test-planets were started with nearly zero initial 
eccentricities and inclinations. The secular frequency $g$ 
was deduced according to the 
secular linear approximation \citep[see e.g.][]{mur-der99}:
\begin{equation} \label{E1}
g =  {\frac{n}{4}}  \sum_{i=1}^{2} {{\frac{m_i}{m_0}} \alpha_{i}^2
b^{(1)}_{3/2}(\alpha_{i})}
\end{equation}
where $\alpha_{i} = {{a}/{a_{i}}}$ for
$i=1,2$ and $a_1, a_2, a$  are the semi-major axes of
the giant planet, the secondary and the test-planet, respectively. $m_1$
and $m_2$ are the masses of the giant planet and the secondary and $m_0$ is the
mass of the primary star. $b^{(1)}_{3/2}$ is a Laplace coefficient.
According to Eq.~\ref{E1}, the value of $g$  depends on the semi-major axes
and the masses, which are all constant for a
certain configuration. Therefore, 
the value of $g$ is also constant for a given semi-major axis of a
test-planet. \\
For calculations of proper frequencies usually a transformation to new
variables\footnote{We do not take into account
  variables associated with the inclination and node (usually called $p,q$ in
  these variables), as we consider planar
  motion.} ($h, k$)  is made, where $h=e \sin \varpi$ and $k= e \cos
\varpi$. Then, the general solution of the test-planets' motion is given by: 
\begin{equation} \label{E2}
\begin{split}
h(t) = e_{free} \sin (g t + \Phi) - \sum_{i=1}^{2} {{\frac{\nu_i}{g-g_i}} \sin
(g_i t + \Phi_i)} \\
k(t) = e_{free} \cos (g t + \Phi) - \sum_{i=1}^{2} {{\frac{\nu_i}{g-g_i}} \cos
(g_i t + \Phi_i)} 
\end{split}
\end{equation}

\noindent where $e_{free}$ and $\Phi$ are constants given by the initial
conditions. The
second part of the two Eqs.~\ref{E2} varies with time as it depends on the
secular 
solution. A secular resonance occurs in case $g_i \approx g$. In such a
case,  the forced
eccentricity will increase rapidly to 1 leading to an escape or close
encounter with another body of the system (as it can be seen in Fig.~\ref{Fpf}
in the area between 0.38 and 0.43 au. For details about this theory see
e.g.~\cite{mur-der99}.\\

\noindent The analytically determined proper periods of the test-planets
in the area between 0.2 and 0.6 au  are
shown by the red dots in the lower panel of Fig.~\ref{Fpf}. 
A comparison with the black dots in this figure shows the good agreement with 
the numerical values derived from the Fourier analysis of
the orbits. 
Especially in the region outside the secular resonance ($a_{TP} > 0.43 au$) we
have found perfect 
conformity of the two results. While for the area closer to the host-star
the proper periods of the analytical solution are slightly larger. 

\subsection{The location of the linear SR -- a new
  semi-analytical method}

\begin{figure} 
\centering
\includegraphics[scale=.9]{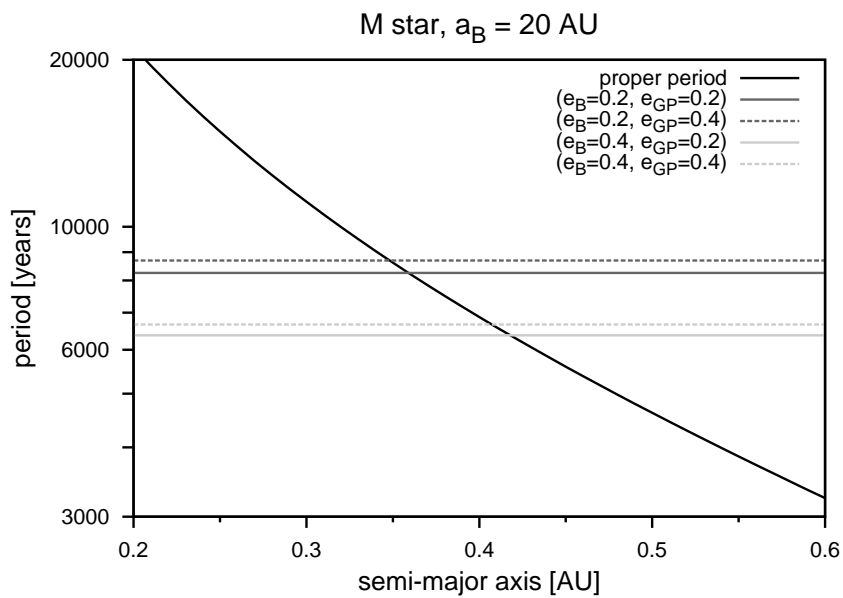} 
\caption{The semi-analytical method applied to the HD41004 binary using
  different eccentricities (0.2 and 0.4) for the giant planet and for the
  binary as shown by the legend. The black curve represents proper periods of
  the test-planet which were analytically determined. The different horizontal
lines show the proper period of the giant planet in the various binary
configurations. The location $a_{SR}$ is defined by the intersection of a
horizontal line with the black curve.}
\label{Fpf2}
\end{figure}
As the analytical solution
represents a good approximation for the proper periods of the test-planets,
the position of the SR (= a$_{SR}$)  can be determined from the intersection of
the red and the blue lines in the lower panel of Fig.~\ref{Fpf}.
Consequently, we need only one numerical computation of the dynamical
  system in order to determine a$_{SR}$. The application of this
semi-analytical method to the HD41004 binary using different eccentricities
for the binary and the giant planet is shown in 
Fig.~\ref{Fpf2} where one can recognize a strong shift of $a_{SR}$
for a change of $e_B$ from 0.2 to 0.4,  which is visible by a comparison of
either the two (horizontal) full lines or the two dashed lines. The different
grey shades 
indicate different eccentricities of the binary ($e_B=0.2$ see the dark grey
lines  and $e_B =0.4$ the light grey lines). While different eccentricities of
the giant  
planet are shown by different line styles (the full line represents the result
for $e_{GP}=0.2$ and the dashed line for $e_{GP}=0.4$).
A comparison of the proper periods for two different $e_{GP}$'s
of a certain $e_B$ shows only a small shift in $a_{SR}$. \\
The intersection of the curve representing the analytical solution for the
proper periods of the test-planets (i.e.\ the black curve in Fig.~\ref{Fpf2})
with the numerically 
determined proper period of the giant planet (i.e.\ the horizontal lines for
the different $e_B$ and $e_{GP}$ in
Fig.~\ref{Fpf2}) defines the position of the linear SR
but not the width which can be quite large in case
of high eccentric motion of the giant planet (as shown in
Fig.~\ref{F3} right panel). The vertical white line indicated the location of
the SR. 

\begin{figure} 
\centering
\includegraphics[scale=.9]{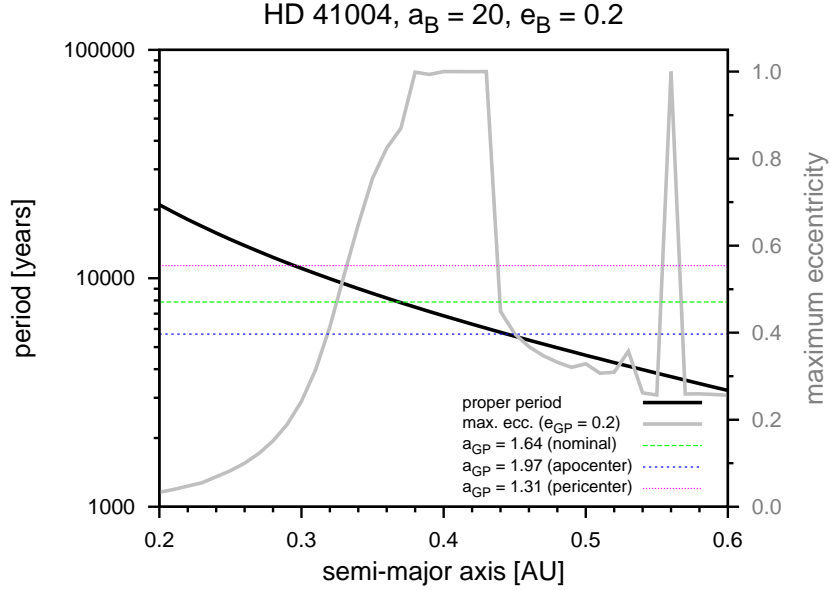} 
\caption{The application of the semi-analytical method to the apo- and
  peri-center positions of the giant planet in the HD41004 binary. The proper
  period of the giant planet derived for these positions are shown by the blue
  and magenta lines for the apo-center and peri-center, respectively. The green
  line labels the proper period for the giant planet for $a_{GP}$. The black
  curve shows the proper periods of the test-planets and the light grey curve
  marks the maximum eccentricities of the test-planets. The intersection of
 the black curve (i) with the green line indicates $a_{SR}$; (ii) with the
 blue line marks the outer border of the linear secular resonance and its
 inner border is defined by (iii) the intersection 
 with the magenta line. For more details see the text.}
\label{Fpf3}
\end{figure}

One can notice that this location is not
centered in the V-shape of the perturbed area.
 It is more or less the inner boundary of the perturbed area. Moreover, we
recognized that the location $a_{SR}$ will not change significantly when
increasing the 
giant planet's eccentricity. Only the width will increase as indicated by the
V-shape in the right panel of  Fig.~\ref{F3}. The two plots show that  the
eccentricity of the binary is important for the location of
the SR while the eccentricity of the giant planet is
responsible for the extension of the SR. 
To determine the width of the SR, we replaced $a_{GP}$ by the peri- and
apo-center distances\footnote{For $e_B$ and $e_{GP}=0.2$: the apo-center is at
  1.97 au and the peri-center is at 1.31 au} of the giant planet 
and calculated the 
proper frequency of the giant planet for these positions. The result is shown
in Fig.~\ref{Fpf3} by the horizonal lines where the green line labels the
proper period for $a_{GP}$, the blue line shows the result for
$a=a_{apocenter}$ and the magenta line 
represents the result for $a=a_{pericenter}$. Looking at the intersections of
these lines with the black curve (representing the proper periods of the
test-planets) one can see from the crossing point of the blue line (at 0.445 au)
that the
apo-center position defines well the outer rim of  the
SR. This is confirmed by the grey line representing the maximum
eccentricities of the test-planets  which shows a sudden change  in max-e in
this area. However, the intersection point 
of the magenta line is not recognized immediately as inner border of
the SR. Only if we examine this positions (i.e.\ 0.295
au) more closely  we notice that at this intersection the max-e value starts to be larger than  $e_{GP}$  and it 
 grows smoothly towards 1 when approaching $a_{SR}$.
The resulting shape of max-e curve 
seems to be significant for regions where SRs are acting, as this shape
was also found in a study by \cite{malho98}. \\
The application of our method to the apo- and
peri-center positions of 
the giant planet determined obviously quite well the width of
the SR (at least for an eccentricity of 0.2 for the
binary and the giant planet).

\subsection{A study of different binary configurations of HD41004}

The successful application of our newly developed semi-analytical method 
motivated us to study the different binary-planet configurations of
  HD4104\,AB of section 3 again with the aid of this method
 to verify if we obtain similar
features as in our numerical investigation. Therefore, we varied (i) the
semi-major 
axis of the binary from 10 au to 40 au, (ii) the eccentricities of the
binary and of the giant planet using either 0.2 or 0.4, and (iii)
the mass of the secondary star (using 0.4, 0.7 and 1 solar mass).
For the resulting 48 
binary-planet configurations we determined $a_{SR}$ applying our  method. 
\begin{table}
  \begin{center}
  \caption{Location of the SR ($a_{SR}$) for different HD41004 binary configurations:}
  \vspace{5mm}
  \label{tab3}
{\scriptsize
  \begin{tabular}{|c||c|c||c|c|}\hline 
 & \multicolumn{2}{|c|}{$e_B = 0.2$} & \multicolumn{2}{||c|}{$e_B = 0.4$} \\
\hline
$a_B$ [au] & $e_{GP} = 0.2 $ &  $e_{GP} = 0.4 $ &  $e_{GP} = 0.2 $ &  $e_{GP}
    = 0.4 $ \\ 
\hline
\multicolumn{5}{|c|}{$a_{SR}$ [AU] for a M-type star secondary:}\\
\hline
10 & 0.96 & 0.94 & 1.08 & -- \\
20 & 0.36 & 0.35 & 0.42 & 0.41 \\
30 & 0.17 & 0.16 & 0.20 & 0.19 \\
40 & 0.10 & 0.09 & 0.11 & 0.11 \\
\hline
\multicolumn{5}{|c|}{$a_{SR}$ [AU] for a K-type star secondary:}\\
\hline
10 & 1.07 & 1.08 & -- & -- \\
20 & 0.50 & 0.48 & 0.58 & 0.56 \\
30 & 0.24 & 0.23 & 0.28 & 0.27 \\
40 & 0.14 & 0.13 & 0.16 & 0.15 \\
\hline
 \multicolumn{5}{|c|}{$a_{SR}$ [AU] for a G-type star secondary:} \\
\hline
10 & 1.25 & 1.25 & -- & --\\
20 & 0.60 & 0.58 & 0.68 & 0.67 \\
30 & 0.30 & 0.29 & 0.35 & 0.34 \\ 
40 & 0.18 & 0.17 & 0.21 & 0.20 \\
\hline
\hline
 \end{tabular}
}
 \end{center}
\vspace{1mm}
\end{table} 
The results of this study are summarized in table \ref{tab3}. This
  overview shows clearly that an increase of the secondary's mass from 0.4
  (M-type star) to 1 solar mass (G-type star) causes an outward shift of the SR 
 to larger semi-major axes for a binary configuration.
 While an increase of the distance between the two stars (here from 10 to
40 au) indicates an inward shift of the SR. As already pointed out in the
  previous section, table \ref{tab3} shows clearly that a change 
of $e_B$ leads to a stronger displacement of the secular resonance than a
change in $e_{GP}$. A comparison with the numerical computations of the previous
sections shows that the results of both studies are in good agreement which
demonstrates the accurateness of this method.

\section{Conclusion}

We studied the planetary motion around one stellar
component of a binary system. As dynamical model we used binary configurations
resembling the HD41004\,AB system where a planet (HD41004\,Ab) has been
detected at 1.64 au \citep{zuck+03,zuck+04}. The aim of this work was to continue and
improve  the study of \cite{pilo05} and to figure out in detail the influence
of a secondary star on the motion of test-planets in the area between 0.15 and
1.3 au.  \\
First we showed in a purely numerical approach the perturbations due to the
giant planet and the secondary star. The max-e plots display the MMRs
and SRs where the latter changed place when we varied the
orbital parameters of the binary. A similar behavior has been found by
\cite{pilo05} for a variation of the giant planet's semi-major axis. 
Depending on the binary-planet configuration the secular resonance can
influence also the motion in the habitable zone. Moreover, our
system configurations showed that test-planets with semi-major axes  $a <
a_{SR}$ (i.e. the semi-major axis of the secular resonance) move on nearly
circular orbits and provide therefore, best conditions for habitability from
the dynamical point of view. \\
In the second part of this investigation,  we developed a
semi-analytical method which allows a fast determination of
$a_{SR}$. Only one numerical integration of the binary-planet
configuration is needed  to determine the proper frequency of the
giant planet. While the test-planets' proper frequencies were taken from the
  analytical solution  of the
Laplace-Lagrange secular perturbation theory. The intersection of both
results defines the location of the SR. \\
We showed that the results for the binary HD41004\,AB of a purely numerical
study and of our semi-analytical approach are in good agreement 
Further applications of this method to tight binary star systems 
with a detected  planetary companions in circumstellar motion are shown 
by \cite{baz+16}.

\begin{acknowledgements}   

\noindent The authors want to acknowledge the Austrian Science Fund (FWF) for
the financial support of this work which was carried out in the framework of
the projects P22603-N16 and S11608-N16 (a sub-project of the NFN Project
``Pathways to Habitability''). EP-L wants to thank Dr.\ P.\ Robutel from 
IMCCE (Paris, France) for fruitful discussions about frequency analysis of
planetary motion. Finally, we thank the unknown referee for helpful
suggestions to improve the paper.

\end{acknowledgements}

\end{document}